\documentclass[aps,prb,twocolumn]{revtex4}
\usepackage{graphicx}\usepackage{color}
\begin{document}

\newcommand{\figureheight}{8.2 cm}
\newcommand{\putfig}[2]{\begin{figure}[h]
        \special{isoscale #1.bmp, \the\hsize \figureheight}
        \vspace{\figureheight}
        \caption{#2}
        \label{fig:#1}
        \end{figure}}

\newcommand{\eqn}[1]{(\ref{#1})}

\newcommand{\be}{\begin{equation}}
\newcommand{\ee}{\end{equation}}
\newcommand{\bea}{\begin{eqnarray}}
\newcommand{\eea}{\end{eqnarray}}
\newcommand{\bean}{\begin{eqnarray*}}
\newcommand{\eean}{\end{eqnarray*}}

\newcommand{\nn}{\nonumber}




\title{ Electronic screening and correlated superconductivity in  carbon nanotubes}
\author{{S. Bellucci}$^1$, M. Cini$^{1,2}$,  P. Onorato$^{1,3}$ and E. Perfetto$^{4}$ \\}
\address{
        $^1$INFN, Laboratori Nazionali di Frascati, P.O. Box 13, 00044 Frascati, Italy \\
        $^2$Dipartimento di Fisica,  Universit\`{a} di Roma Tor Vergata, Via della Ricerca Scientifica 1 00133,  Roma, Italy\\
$^3$Department of Physics "A. Volta", University of Pavia, Via Bassi 6, I-27100 Pavia, Italy\\
$^4$Consorzio Nazionale Interuniversitario per Le Scienze Fisiche
della Materia, Universit\`{a} di Roma Tor Vergata, Via della
Ricerca Scientifica 1 00133,  Roma, Italy}
\date{\today}

\begin{abstract}
A  theoretical analysis  of  the  superconductivity
observed
 recently in Carbon nanotubes is  proposed. We argue that  ultra-small  (diameter $ \sim
0.4 nm$) single wall carbon nanotubes (with transition temperature
$T_c\sim 15\; ^{o}K$)  and  entirely end-bonded multi-walled ones (
$T_c\sim 12 \;^{o}K$) can superconduct  by an   electronic
mechanism, basically the same in both cases. By a Luttinger liquid
-like approach, one finds enhanced superconducting correlations due
to the strong screening of the long-range part of the Coulomb
repulsion. Based on this finding, we perform a detailed analysis on
the resulting
      Hubbard-like model, and calculate   transition temperatures of the same
order of magnitude as the measured ones.

\end{abstract}

\maketitle

\section{Introduction}

Carbon nanotubes as one-dimensional (1D) molecular conductors are
among the best candidates for investigating the possibility of 1D
superconductivity. They are basically rolled up sheets of graphene
forming tubes that are only nanometers in diameter and length up to
some microns. Carbon nanotubes may be grown as single-walled (SWNT)
or multi-walled (MWNTs), that are typically made of several
coaxial graphene shells. They show semiconducting
or metallic properties depending on the helicity of the carbon rings
around the tubule\cite{saito}.

The electron-electron interactions are also known to modify
significantly the transport properties of the nanotubes\cite{yao},
leading to the breakdown of the conventional Fermi liquid picture.
In fact the 1D character of the system leads to a strong correlation
among electrons, inducing of  the so-called Luttinger
liquid\cite{lutt}. The Luttinger liquid behavior is characterized by
a power-law suppression of physical observables, such as the tunnelling
conductance, over a wide range of temperatures. Indeed the tunneling
conductance $G$ reflects the power law dependence of the
tunneling density of states (DOS), in  a small bias
experiment\cite{kf}
 \bea G=dI/dV\propto T^{\alpha} \eea
for $eV \ll k T$, where $V$ is the bias voltage, $T$ is the
temperature and $k $ is Boltzmann's constant. Evidence of
Luttinger liquid behavior in SWNTs  has been found in many
experiments\cite{ll,ll2,17}, where the temperature dependence of the
resistance above a crossover temperature $T_c$ was
measured\cite{Fischer}. The critical exponent $\alpha$ assumes
different values   for an electrode-bulk junction ($\alpha_{bulk}$)
and for an electrode-end junction ($\alpha_{end}$), as  reported for
MWNTs in Ref.\onlinecite{1b}.

Experiments have been also carried out to show  superconducting
correlations in SWNTs at low temperatures. Clear
evidence of superconductivity was found in nanotubes suspended
between superconducting contacts, showing the so-called proximity
effect\cite{7,8}. Moreover, genuine superconducting transitions
below $1K$ have been observed in thick ropes of nanotubes suspended
between normal and highly transparent electrodes\cite{10b}.

A few years ago, ultra-small-diameter single wall
nanotubes (USCN), with diameter of $\sim 0.4\; nm$, have been
produced  inside the channels of a zeolite matrix. Possible metallic
geometries compatible with such a small radius are the armchair
(3,3) and the zig-zag (5,0) ones. The ultra-small diameter of these
tubes induces many unusual properties, such as a superconducting
transition temperature $T_c \approx 15 K$\cite{[11]}, much larger
than that observed in bundles of larger diameter tubes\cite{ropes}.

Quite recently\cite{tk} a similar transition
temperature was observed in entirely end-bonded MWNTs. It was found
that the emergence of superconductivity ($T_c=12K$) is highly
sensitive to the junction structures of the Au electrode/MWNTs.

The question arises, whether the superconductivity
in the MWNTs\cite{tk} can be understood on the same grounds as the
superconductivity of the USNTs\cite{noijpcm}. Here,
  we argue that a purely electronic mechanism could work in both cases.
We start by a Tomonaga-Luttinger model of the electronic system,
focussing on the most relevant sources of screening of the Coulomb
repulsion. The long-range part of the interaction can be strongly
reduced due to the peculiarity of the experimental conditions. This
opens up the possibility of a breakdown of the Luttinger liquid
regime towards a pairing instabilty. Anyway this finding calls for a
more detailed analysis based on  an effective Hubbard model,
emphasizing the role of the short range interaction.
We have in mind a basically electronic mechanism,
but  lattice effects should be considered as well. The analysis
could be carried out along the lines of Ref. \cite{fononi}, where
the interplay of phonons with the W=0 mechanism was discussed. Since
this is very demanding in the present geometry the problem is
deferred to future publications.

\section{Luttinger liquid in carbon nanotubes}

The Luttinger liquid is the prototype of interacting electrons in 1D
and it is governed by the so called Tomonaga-Luttinger Hamiltonian.
In this model, the electrons have linear dispersion relation around
positive (Right) and negative (Left) Fermi points located at $\pm
K_{F}$ and the e-e interactions act only between Right
or Left electron densities. This means that only
the {\it forward scattering} component with small momentum transfer
$q \sim 2 \pi/L \equiv q_c$ of the Coulomb repulsion (denoted by
$g_2$) is retained, while the {\it backscattering} component with $q
\sim 2K_F$ (denoted by $g_1$) is assumed to be negligible. Anyway,
even a small $g_1$ is important at low temperatures
since it may induce the breakdown of the Luttinger liquid state
toward a phase transition.

In normal conditions, a carbon nanotube is composed
itself by two coupled (identical) Luttinger liquids, since there are
a Left and a Right linear branch respectively around each of the two Fermi
points  at $(\pm
K_F,0)$ ($K_F=4\pi/3a$, and $a=2.46${\AA} is the lattice constant).
These branches are highly linear with
Fermi velocity $v_F\approx 8\times 10^5$ m/s. The linear dispersion
relation holds for energy scales $E < D$, with the bandwidth cutoff
scale $D\approx \hbar v_F/R$ for tube radius $R$.


In order to study the effects of interactions, we  introduce the unscreened Coulomb
repulsion in a wrapped 2D geometry
 \bea \label{U}
 V_0(x-x',y-y')=\frac{e^2/\kappa}{\sqrt{(x-x')^2+4
R^2 \sin^2(\frac{y-y'}{2R})}}, \eea
where $x$ denotes the coordinate along the tube axis and $0<y<2\pi R$ is the coordinate along the
circrcumference of the transverse cross-section of the tube.
The Fourier transform  $\hat{V}_0(q)$ reads
 \bea\label{uq}
 \hat{V}_0(q)\approx \frac{e^2/\kappa}{\sqrt{2}}  \left[K_0(\frac{qR}{2})I_0(\frac{qR}{2}) \right],
\label{uq2} \eea
 where  $\kappa$ is the dielectric constant, $K_0(q)$   denotes the modified Bessel
function of the second kind, $I_0(q)$ is  the modified Bessel
function of the first kind. The nanotube radius $R$ yields a
natural cutoff $\approx \frac{2\pi}{R}$ of the interaction.

According to the above discussion we have $g_2= \hat{V}_0(q_c)$ and $g_1 = \hat{V}_0(2 K_F)$.
Following Ref.\onlinecite{lutt}, we introduce an additional interaction
($f$) which measures the difference between intra- and
inter-sublattice interactions.
Such a term is not contained in Eq.(\ref{uq}) and it is due to the hard core of the
Coulomb interaction.

The Luttinger liquid behavior in carbon nanotubes  was theoretically
investigated in Ref.\onlinecite{lutt}, where the low-energy theory
including Coulomb interaction is derived. Although the analysis is
quite general, explicit results were obtained for typical
metallic nanotubes, i.e. armchair $(10,10)$ SWNTs  with radius $R
\approx 1.4nm$ and length $L \approx 3\mu m$, which  we name
$CN_{10}$.

The Luttinger parameter $g$ depends just on the forward scattering part
of the interaction
\bea\label{K}
\frac{1}{g}= \sqrt{1 + \frac{g_2}{ (2
\pi  {v}_F)}} , \eea
whereas the critical exponent can
be written in terms of $g$ as $
  \alpha_{bulk}=\frac{1}{4} \left(g+ 1/g-2 \right)$.
A detailed estimate for the $CN_{10}$ gives $g\approx 0.2$ and $\alpha_{bulk}\approx 0.32$, in
agreement with experiments\cite{lutt}.

Concerning the rest of the couplings, it is shown that both $g_1$ and $f$  scale as $1/R$ and in $CN_{10}$ they are much smaller than $g_2$\cite{lutt}.
Anyway at low temperature their effects  should be included. In Ref.\onlinecite{lutt} this
has been realized  by means of a renormalization group calculation.

The main result is the existence of two different crossover temperatures, namely
$k T_f=  D e^{-\frac{2\pi v_F}{f}}$ and
$k T_b=  D e^{-\frac{2\pi v_F}{g_1}}$ associated to the dominance of $f$ and $g_1$ respectively.
Below these temperatures the Luttinger liquid breaks down and a (quasi-) long-range
order phase appears.
For long-ranged interactions (which is the case of nanotubes in typical conditions),
we have $T_f \sim T_b$, while for short-ranged interactions it results $T_f < T_b$. In the latter case a superconducting instabilty
is predicted at $T \sim T_f$ if the Luttinger liquid parameter $g$ is larger than $1/2$. This condition implies a very strong screening of $g_2$ and it is the main aim of the present paper
to show that such an instance can be realized in the experimental conditions of Refs.\onlinecite{[11]} and \onlinecite{tk}.

Anyway it is worth to note that the transition temperature for $CN_{10}$ with the typical long-range interaction
as in Eq.(\ref{U}) is estimated to be $T_f \sim T_b \sim 1mK$,
i.e. a value certainly hard to be observed. This is due to the
smallness of $g_1$ and $f$ (compared to $g_2$), since they scale as $1/R$ and are
sizeable only for very thin tubes.


Starting from the results of Ref.\onlinecite{lutt} it appears that a pure
electronic mechanism consistent with an observable superconductivity in the
nanotubes requires a strong reduction of the forward scattering $g_2$ (so that $g>1/2$) and the increase of  $g_1$
and  $f$ with respect to the values of typical samples.

\section{Ultra small nanotubes}

In this Section we analyze the screening properties of ultra small  nanotubes,
due to the experimental setup of Ref.\onlinecite{[11]}.
In particular we show that the presence of many nanotubes inside the zeolite matrix provides
 a strong screening of the long range component of the electron-electron
 interaction ($g_2$), while the short range components have
to remain almost unchanged. This allows for the
occurrence of a sizable superconducting instability within the
Luttinger liquid approach. In what follows we focus on the (3,3)
nanotubes as the main candidate
 to be present inside the zeolite matrix (we shall refer to them as $CN_{3}$),
  although the possibility of a presence of (5,0) zig-zag tubes cannot be discarded.

As already pointed out in Refs.\onlinecite{gonzperf} and
\onlinecite{gonzperf2}, the intra-tube Coulomb repulsion at small
momentum transfer (i.e. in the forward scattering channel) is
efficiently screened by the presence of electronic currents in
neighboring nanotubes. In the experimental samples of
Ref.\onlinecite{[11]} the carbon nanotubes are arranged in large
arrays with triangular geometry, behaving as a genuine 3D system. By
means of a generalized Random Phase Approximation (RPA) approach, it
is shown\cite{gonzperf},\cite{gonzperf2} that the forward scattering
parameter $g_{2}$ gets renormalized according to
\begin{equation}
g_{2}^{CN_{3}} \rightarrow (\hat{U})_{s,s} \, ,
\label{}
\end{equation}
where $ (\hat{U})_{s,s'}$ obeys
to the Dyson equation
\begin{equation}
(\hat{U})_{s,s'}=[\hat{U}_0(1-\hat{U}_0 \Pi_0)^{-1}]_{s,s'} \,
\label{screened}
\end{equation}
where $\Pi_{0}= 1/2\pi v_{F}$ and the matrix $(\hat{U}_0)_{s,s'}$ is given by
\begin{equation}
(\hat{U_0})_{s,s'} = \frac{1}{3^2}\int_{0}^{L}dx \sum_{y,y'}
U_{0}(x,y,y',s,s')
 \, ,
\label{}
\end{equation}
Here $U_{0}(x,y,y',s,s')$ is the 3D bare Coulomb repulsion between two electrons in $(3,3)$
nanotubes in the zeolite matrix such that $x$ is the relative coorditate in the longitudinal
direction, $s, s'$ indicate the position of the two nanotube axes in the zeolite matrix and
$y,y'$ are the circular coordinates in each nanotube.
We note that if $s=s'$ we recover $U_{0}(x,y,y',s,s)=V_0$ as in Eq.(\ref{U}).

The above
source of screening provides $g_{2}^{CN_{3}}/v_F =(\hat{U})_{s,s}/ v_F \approx 0.7$ (see Fig.5 of Ref.\onlinecite{gonzperf2}), where it is used that the distance between nearest
neighbors nanotubes in the matrix is 1nm.  This result indicates
a large reduction (by a
factor $\approx 10^{-2}$) with respect to the bare coupling. The backscattering coupling
$g_{1}$ is not affected appreciably.


Let us analyze now how the couplings $g_1$ and $f$ are modified in
USNTs. As shown in Ref.\onlinecite{lutt}, they scale as $1/R$ and
therefore are $10/3$ times larger in the  $CN_{3}$ than in the
$CN_{10}$. The additional forward scattering $f$ corresponds
to\cite{lutt} $\delta V_p= V_{++}-V_{+-}$, where $V_{p,p'}$ is the
interaction between electrons belonging to different sublattices
($p,p'$), and it is strongly suppressed at a distance much larger than
$\ell\sim 0.3 nm$\cite{noijpcm}. In the same way, the only
non-vanishing contribution to $g_1$ comes from $|x-x'|\leq a$,
because of rapidly oscillating contributions\cite{lutt}.

As a consequence we find the relation $g_1^{CN_{3}} \sim
\frac{10}{3} g_1^{CN_{10}}$ and the same also holds for $f$. All
of the above mentioned effects have a strong impact on the
crossover
 temperature $T_c$, and following ref. \cite{lutt}  we
 estimate
\bea
 \label{T_B} k T_b \sim kT_f  \sim k T_c \propto D e^{-\frac{2\pi v_F}{g_1}}.
 \eea
For the $CN_{10}$,  $T_c$ was estimated as $\sim 0.1 mK$, or some
order of magnitude larger for well-screened interaction\cite{lutt}.
It follows that in USNTs $T_c$ should be several orders of magnitude
larger than the one predicted for a $CN_{10}$ with a factor
compatible with the observed critical temperature.\cite{nota} To sum
up,  one can expect that the screening of the short-range part of
the Coulomb interaction  be less efficient in CNTs than in bulk
graphite; moreover, in $CN_{3}$  it will be less efficient than in
$CN_{10}$. This is an unavoidable consequence of the reduced
effective dimensionality, which puts constraints on the screening
cloud. We wish to emphasize that the above arguments lead to the
conclusion that the increased short range repulsion does not impair
$T_c$, as one could naively expect; actually the critical
temperature turns out to increase, at least for a moderate increase
of the short-range component of the Coulomb repulsion. This
seemingly paradoxical conclusion is further validated by a thorough
analysis within the Hubbard model, as shown in the next subsection.

\subsection{Hubbard model for ultra small nanotubes}

The results of the previous Section suggest the possibility of a
superconducting instability in USNTs within the Luttinger liquid
scenario.  Indeed lattice effects and very short range
interactions become dominant so that  the Luttinger liquid picture can break
down at a sizable energy scale. Anyway we point out that, as long as
$g_1$ and $f$  become comparable to $g_2$, all of them
should be treated on the same footing. This indicates that the
system under consideration should be better described in the
Hubbard-like framework, which emphasizes the role of the lattice and the
short range interaction. We recall that the Hubbard Hamiltonian reads:
\begin{eqnarray}
H&=&H_{0}+W \nonumber \\
&=&t\sum \limits_{ \langle {\bf r},{\bf r}' \rangle }\sum
 \limits_{\sigma}\left(c^{\dagger}_{{\bf r},\sigma}c_{{\bf r}',\sigma}+h.c. \right)
+U\sum \limits_{{\bf r}} n_{{\bf r},\uparrow }
n_{{\bf r},\downarrow  } ,
\label{hub}
\end{eqnarray}
where $c^{\dagger}_{{\bf r},\sigma}$ ($c_{{\bf r},\sigma}$) is the creation
(annihilation) operator of a graphitic $p_{z}$ electron of spin $\sigma$ on the wrapped honeycomb
lattice  site
${\bf r}$, the sum runs over the pairs $\langle {\bf r},{\bf r}'
\rangle$ of nearest neighbour carbon atoms,  $n_{{\bf r},\sigma}=c^{\dagger}_{{\bf r},\sigma}c_{{\bf r},\sigma}$ is the number operator referred to the site ${\bf r}$, $t$ is the hopping
parameter and $U$ is the on-site Hubbard repulsion.

In Ref.\onlinecite{krot} the superconductivity in carbon nanotubes described by
Eq.(\ref{hub}) was investigated with the renormalization group technique.
Unfortunately this approach does not allow for a stringent
prediction of the critical temperature.

In Ref.\onlinecite{psc} it was proposed an electronic mechanism
which leads to superconducting pairing starting from the Hubbard
model on the wrapped honeycomb lattice  away from half filling. In
this approach all the interaction channels are considered on the
same footing and this makes it possible to predict
a reliable transition temperature  for the USNTs\cite{noijpcm}.

The findings of Ref.\onlinecite{psc} are based on the so called
$W=0$ theory\cite{psc2}; this provides a singlet pairing mechanism
operating on a lattice with Hubbard interaction and is otherwise
somewhat  analogue to  the Kohn-Luttinger mechanism\cite{kohn}. On
the basis of symmetry arguments, it is possible to show that the
Hamiltonian in Eq.(\ref{hub}) admits two-body singlet eigenstates
with no double occupancy on the honeycomb sites, called  $W=0$
pairs. $W=0$ pairs are therefore  eigenstates of the kinetic
energy operator $H_{0}$ and of the Hubbard repulsion $W$ with
vanishing eigenvalue of the latter. As a consequence the electrons
forming a $W=0$ pair have no direct interaction and are good
candidates to achieve bound states. As a consequence the electrons
forming a $W=0$ pair have no direct interaction and are good
candidates to achieve bound states. Their effective interaction
$V$\cite{psc} comes out from virtual electron-hole  excitation
exchange with the Fermi sea and in principle can be
attractive\cite{psc2}. The binding energy $\Delta$ of the $W=0$
pairs for an armchair $(n,n)$ nanotube can be obtained by solving
the gap equation
\begin{equation}
\frac{1}{V}=\frac{1}{8n}\frac{1}{2\pi} \sum_{k_{y}} \int  dk_{x}
\frac{ \theta \left(\varepsilon(k_{x},k_{y}) -\varepsilon_{F}
\right)} {2(\varepsilon(k_{x},k_{y} ) -\varepsilon_{F})-\Delta} \,
, \label{asym}
\end{equation}
where $\varepsilon_{F}$ is the Fermi energy,
$\varepsilon(k_{x},k_{y})$ is the eigenvalues of $H_0$ relative to
momentum $(k_{x},k_{y})$ ($x$ is along the tube axis and $y$
denotes the transverse direction). We recall that the effective
interaction $V$ between the two electrons forming the $W=0$ pair
encodes an indirect interaction mediated by the exchange of
virtual electron-hole excitations. $V$ is a complicated function
of the Hubbard $U$, but does not contain any $o(U)$ contribution
because of the $W=0$ property\cite{psc}.

Remarkably it is found\cite{psc} that $\Delta$ is nonvanishing
only for the doped systems and   increases with decrasing $R$ and
with increasing $U$, at least for moderate repulsion. Such a
result confirms that in USNTs the superconducting phase is
supported by the small size (i.e. larger Hubbard repulsion) in the
presence of doping. Indeed, the $W=0$ mechanism requires moving
the Fermi level away from half filing. This can be achieved in
several ways.  CNTs  can exchange charge with the surroundings by
doping or by contact potential difference, due to the contacts
with the zeolite matrix and/or the electrodes. The innermost
nanotubes in multiwalled structures  can conceivably exchange some
charge with the outer ones.   A shift of the order of $10^{-1}$eV
is sufficient to produce a sizable $\Delta$ in very small CNTs
(see Fig. 5 in Ref.\onlinecite{psc}). Note that it would be hard
to ensure perfect neutrality of such tiny structures in
inhomogeneous environments, so it is reasonable to assume that
such a small shift in the Fermi level is easily realized.

Now, the first step in order to have a quantitative estimate of
$T_c$ in USNTs is a sensible evaluation of $U$. Because of the
screening property due to the experimetal conditions we find
$g_{2}^{CN_{3}} \sim g_{1}^{CN_{3}} \sim f^{CN_{3}} \sim U a/n$.
Therefore $g_{2}^{CN_{3}}/v_{F}\approx 0.7$ implies  $U \approx
0.7\times 3 \times v_{F}/a \approx 4.2$eV. In Ref.\onlinecite{psc}
$U/t =1.6$ was used (with $t=2.6$eV the hopping parameter of
graphitic honeycomb lattice), which means $U=4.4$eV. This is
reasonable in the  light of other data available. In bulk
Graphite, the Auger line shape analysis\cite{cida} gives a
repulsion $U=5.5$ eV between $p$ holes. Therefore, the results of
Ref.\onlinecite{psc} fairly apply to the case under consideration
and we can extrapolate $\Delta \approx 8$meV for (3,3) nanotubes
at optimal doping. Finally, the BCS formula $\Delta = 1.76 kT_c$
for the mean field transition temperature gives
$$
T_c\approx 7 \div 70 K
$$
which is compatible with the measured one.
We observe that the lower boundary $T_c \sim 7K$ takes into account that
$\Delta$ may vary of about one order of magnitude away from optimal doping.

Finally we remark that the corresponding
$T_c$ for the $CN_{10}$ is of the order of the $mK$ in agreement
with the predictions of Ref.\onlinecite{lutt}.


%
%

\section{Entirely end-bonded MWNTs}
 The experiment of Ref.\onlinecite{tk} has shown that
"entirely end-bonded" MWNTs can superconduct at temperatures as high
as $12 \;^{o}K$. In this system the authors claim
that almost all the shells of the MWNT's are electrically active.
Such a high quality of the contacts seems to be crucial, in order to
observe the superconducting transition at such a high temperature.

Moreover the clear power-law of the conductance observed for $T>T_c$
is consistent with the Luttinger liquid character of the normal state.
Therefore the observed sharp breakdown of the power-law at $T_c$ is
an indication that our approach based on the superconducting
 instability of the Luttinger liquid is well posed.

Now we discuss some relevant physical consequences of the activation
of several shells. In a typical transport experiment, only the
outermost shell of the MWNT becomes electrically active. As a
consequence the conducting channel is not efficiently screened and
retains a strong 1D character. On the other hand, the activation of
the internal shells gives a large dielectric effect, due to intra-
and inter-shell screening, and at the same time it provides an
incipient 3D character, which is crucial for establishing the
superconducting coherence.

We assume that all contacted shells can transport the normal
current as resistors in parallel connection. Therefore at $T>T_c$,
the electrons flow in each shell. It is however clear that the
conductance $G$ is mainly given by the outermost shells, because they
have more conducting channels due to larger radius.

For what concerns $T<T_c$, we know from the previous Section that
superconductivity is favored in the inner part of the MWNT, where
the radius of the shells is reduced. In particular, we focus our
attention on the innermost shell corresponding to a radius as small as
$R_{in} \approx 0.4n$m (e.g. a (6,6) armchair). We can wonder whether this shell
 can display a superconducting transition, and what is the corresponding $T_c$.

Following the discussion reported above we have to evaluate the
screening of the long range interaction $g_2$ which determines the
forward scattering coupling and the corresponding values of the
short range terms $g_1$ and $f$.


As already discussed above, the $g_2$ interaction is screened by the
electronic currents located in the sorrounding $(n,n)$ shells.
A similar calculation as is Section III leads to the following renormalization of $g_2$
in a given shell $n$.
\begin{equation}
g_{2}^{CN_{n}} \rightarrow (\hat{W})_{n,n} \, ,
\label{}
\end{equation}
where $ (\hat{W})_{n,n'}$ obeys
to the Dyson equation
\begin{equation}
(\hat{W})_{n,n'}=[\hat{W}_0(1-\hat{W}_0 \Pi_0)^{-1}]_{n,n'} \,
\label{}
\end{equation}
where the matrix $(\Pi_{0})_{n,n'}$ reads
\begin{equation}
(\Pi_{0})_{n,n'}= \sum_{i}\frac{1}{2\pi v_{F,n}(i)}
\end{equation}
where the sum runs over all the Fermi points $i$ (with related Fermi velocities $v_{F,n}(i)$)
in the shell $n$.
The matrix $(\hat{W}_0)_{n,n'}$ is given by
\begin{equation}
(\hat{W_0})_{n,n'} = \frac{1}{n n'}\int_{0}^{L}dx \sum_{y,y'}
W_{0}(x,y,y',n,n')
 \, ,
\label{}
\end{equation}
Here $W_{0}(x,y,y',n,n')$ is the 3D bare Coulomb repulsion between two electrons in $(n,n)$
and $(n',n')$ shells, $x$ is the relative coorditate in the longitudinal
direction,
$y,y'$ are the circular coordinates in each shell.
We note that also in this case  we recover $W_{0}(x,y,y',n,n)=V_0$ as in Eq.(\ref{U}).

For the innermost (6,6) shell, where superconducting correlations
are expected to be enhanced,  we find $g_{2}^{CN_{6}}/v_F
=(\hat{W})_{6,6}/ v_F  \approx 0.5$, which again indicates a very
strong screening. Here we used that the innermost shell has a radius
$R_{in}\approx 0.4n$m, while the outermost shell has $R_{out}\approx
5n$m; this implies that the total number of shells in the MWNT is
$N_{shell}\approx 15$ (we recall that the typical intershell
distance is 0.34nm). Moreover we assumed, that because of doping,
all the shells have a metallic character. We
observe that the multiwalled geometry provides  more efficient
(altough of the same order) screening effect with respect to the
array geometry.

Concerning the rescaling of $g_1$ and $f$, they scale as
$1/R$, as we discussed above. In fact  the short range component
of the interaction is not affected appreciably by the surrounding
conducting channels.

Thus the temperature $T_c$ can be obtained from Eq.(\ref{T_B}) by
taking into account the modified values of $g_1$ and $D$ as
$T_c\sim 2-20K$, which corresponds to a transition temperature
in the same range of values as the one observed for MWNTs.

\subsection{Hubbard model for the innermost shell }

At the light of the above discussion, we predict the presence of a
superconducting instability in the innermost shell of the MWNTs of
Ref.\onlinecite{kf}, where the short-range correlation effects become
dominant.
Again we have to give a sensible estimate for the Hubbard repulsion $U$.

The analysis of the screening properties of the experimetal setup
gives $g_{2}^{CN_{6}} \sim g_{1}^{CN_{6}} \sim f^{CN_{6}} $ for the
innermost shell. Therefore $g_{2}^{CN_{6}}/v_{F}\approx 0.5$ implies
$U \approx 0.5\times 6 \times v_{F}/a \approx 6.0$eV. Therfore the
results of Ref.\onlinecite{psc3} can be applied, where $U/t =2.5$
(and hence $U=6.5$eV) was used. This is somewhat
larger than the Graphite value\cite{cida}.  From that reference we
find $\Delta \approx 5.5$meV for (6,6) nanotubes at optimal doping,
which means
$$
T_c\approx 4 \div 40 K \, .
$$
This value is slightly lower than the one of USNTs, in qualitative
agreement with the experimental findings.
Also in this case the lower boundary $T_c \sim 4K$ is understood in terms of
possible deviation from optimal doping.


Let us now comment upon the main result of this Section.
In usual conditions, transport measurements carried out in MWNTs
reflect the electronic properties of the outer shell,
which the electrodes are attached to.
On the other hand, in entirely end-bonded samples
the inner shells are electrically active, with relevant consequences.
In particular the innermost one is able to support the transport
of Cooper pairs below a
temperature consistent with the measured one.

\section{Conclusions}

\begin{figure}
\includegraphics*[width=1.0\linewidth]{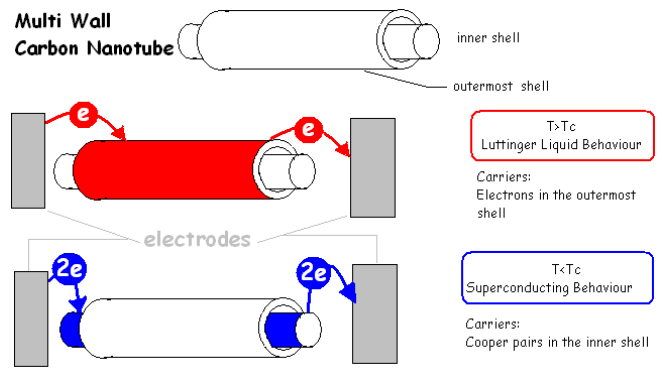}
 \caption{We assume that all shells of an entirely end bonded  MWNT are resistors in
  a parallel connection. For
$T>T_c$ the current is  due to the flow of electrons in the
outermost shells with the typical behaviour of a Luttinger Liquid.
For $T<T_c$ a superconducting transition is allowed in the innermost
shell; thus the transport is due to Cooper pairs.}
\end{figure}

Carbon nanotubes are not naturally superconducting. The main reason
for this is the presence of a stable Luttinger liquid phase, as a
reflection of the strong  electron-electron repulsion, preventing
Cooper pairs to form at sizable temperature.

Some recent experiments have shown that in particular conditions
it is possible to observe superconducting corrrelations, which can compete with
the Luttinger liquid phase and even overcome it.

We propose a scenario where the Luttinger liquid can break down at
sizable energy scales, assuming that (i) the radius of the tube is
small enough, (ii) an efficient screening of the forward scattering
interaction can be achieved. In these conditions a superconducting
instability can arise by a purely electronic mechanism, and a model
based on the Hubbard interaction predicts a crossover  temperature
$T_c$  of the same order of magnitude as the measured one.

In USNTs of Ref.\onlinecite{[11]}, the presence of the many nanotubes in the surrounding
zeolite matrix is quite relevant for the screening of the long range
interaction,
while the small size of the tubes is crucial, in order to increase the
strength of the short range interactions $g_1$ and $f$.
The mechanism  requires doping the nanotube away from half
filling, but a shift of the  Fermi energy by tens of an eV  can produce a
sizable $\Delta$.

In the case of entirely end-bonded MWNTs, the screening of $g_2$
is due to the presence of many shells. Based on this
consideration, we assume that all shells are resistors in a
parallel connection. Therefore the current is mainly due to the
flow of electrons in the outermost shells for $T>T_c$, i.e. in the
Luttinger liquid phase, while the transport of Cooper pairs holds
in the innermost and thinnest shell at $T<T_c$  (see Fig.1). This
scenario is in line with the prediction\cite{psc,psc3} of an
increase in pair binding energy with decreasing nanotube radius.

E. P. was supported by INFN grant 10068.E. P. acknowledges
Consorzio Nazionale Interuniversitario per Le Scienze Fisiche
della Materia for financial support.



\bibliographystyle{prsty} 

\bibliography{}

\begin{thebibliography}{99}
\bibitem{nota}
The analysis of Ref.\onlinecite{lutt}
for a general interaction potential predicts that a crossover
range of temperatures should be observable on
 temperature scales $T_f <  T < T_b$, with a
SC phase just corresponding   to the  regime $T < T_f$ dominated
by the true $T=0$ fixed point. For long-ranged interactions, we
have $T_f\approx T_b$, and the intermediate regime is  not
observable. However in our case we assume a breakdown of the
Luttinger regime below the highest of the two temperatures $T_f$
and $T_c$.
\end{thebibliography}

\end{document}